\documentclass[12pt,preprint]{aastex}

\newcommand{\grad}{\mbox{\boldmath$\nabla$}}

\renewcommand{\v}{{\mathbf{v}}}
\newcommand{\vdot}{{\mathbf{\cdot}}}
\newcommand{\A}{{\mathbf{A}}}
\newcommand{\B}{{\mathbf{B}}}

\newcommand{\vcross}{{\mathbf{\times}}}

\newcommand{\thth}{\hspace{1.5pt}}

\newcommand\Div{\grad\vdot\thth}

\newcommand{\fract}[2]{{\leavevmode\kern.1em
  \raise.5ex\hbox{\the\scriptfont0 #1}\kern-.1em
\kern-.15em\lower.25ex\hbox{\the\scriptfont0 #2}}}

\shortauthors{Gilman \& Dikpati}
\shorttitle{Resonance in forced flux-transport dynamos}

\begin{document}

\title{RESONANCE IN FORCED FLUX TRANSPORT DYNAMOS}

\author{Peter A. Gilman and Mausumi Dikpati}

\affil{High Altitude Observatory, National Center for Atmospheric
Research, 3080 Center Green, Boulder, CO 80307-3000.}

\email{gilman@ucar.edu,dikpati@ucar.edu}

\begin{abstract}

We show that simple 2 and 3-layer flux-transport dynamos, when
forced at the top by a poloidal source term, can produce a widely varying
amplitude of toroidal field at the bottom, depending on how close the 
meridional flow speed of the  bottom layer is to the propagation speed 
of the forcing applied above the top layer, and how close the amplitude of the 
$\alpha$-effect is to two values that give rise to a resonant response. 
This effect should be present in this class of dynamo model no matter how 
many layers are included. This result could have implications for the 
prediction of future solar cycles from the surface magnetic fields of prior 
cycles. It could be looked for in flux-transport dynamos that are more 
realistic for the Sun, done in spherical geometry with differential rotation, 
meridional flow and $\alpha$-effect that vary with latitude and time as 
well as radius. Because of these variations, if resonance occurs, it should
be more localized in time, latitude and radius.

\end{abstract}



\section{INTRODUCTION}

Dikpati et al (2006) first used a flux transport dynamo calibrated to the Sun 
(Dikpati et al 2004) to simulate and predict solar cycle peaks from the record 
of past surface magnetic field patterns. This was done mathematically by 
forcing the dynamo equations at the top boundary, with a forcing function 
derived from past surface magnetic fields. Flux transport dynamos, and indeed 
all dynamos, have their own unforced, usually complex frequencies of excitation
that are commonly found by treating the dynamo equations as an eigenvalue 
problem. Many naturally occurring and man-made systems have such properties.

When a physical system that has natural freqencies is excited by external 
forcing whose own frequency is close to one of the natural ones, there can be
resonance produced--that is, the system will be excited strongly by the forcing
compared to the case where the forcing frequency is not close to a natural one.
The purpose of this paper is to explore the possibility of resonance in 
flux-transport dynamos relevant to the solar cycle.

In flux transport dynamos, there are several physical properties that help
determine the unforced frequencies of the system. These include differential 
rotation, meridional circulation, the so-called $\alpha$-effect, or kinetic 
helicity, and turbulent magnetic diffusion. It is now well established
(Dikpati and Charbonneau, 1999) that unless the magnetic diffusivity is very
large, meridional flow at the bottom of the dynamo layer is primarily 
responsible for the real part of the natural frequency of the dynamo, which 
determines the speed with which induced toroidal and poloidal fields near the 
bottom migrate toward the equator. Therefore the closeness of the frequency of 
forcing at the top to the speed of the flow at the bottom could help determine 
how much dynamo response there is. 

Since the forcing at the top is created by emergence of concentrated magnetic 
flux from the bottom, in the form of active regions, and the rate of movement 
of the zone where active regions are found moves toward the equator (not 
coincidentally) at a rate close to the meridional flow speed near the bottom,
we might expect the conditions for resonance to occur in the bottom layer 
to be favorable. On the other hand, we know from observations (Ulrich, 2010 and
references therein) that the meridional flow at the top of the convection zone 
is toward the poles, opposite to the propagation of the surface forcing as 
well as 5-10 times faster. Thus we should not expect resonance to occur near 
the surface. 

It is also well known (Ulrich 2010 and references therein) that the meridional 
circulation varies with time. This time variation is now being incorporated 
into a flux-transport dynamo used for prediction by Dikpati and colleagues. 
In the 2006 prediction, meridional circulation generally was kept fixed in time.
Dikpati et al (2006), Dikpati and Gilman (2006) recognized that such time 
variations could be important, but felt they lacked sufficient knowledge of its variations to include them. They adjusted the time-independent meridional 
flow amplitude to give the average period of the past solar cycles, and 
stretched or compressed all the surface forcing data to the same period, 
to avoid any artificial or non-physical mismatches between the natural 
dynamo period and the period of the forcing. 

But there can also in principle in the Sun be real differences between the 
period of the top forcing that was created by the previous cycle, and the 
freqency of equatorward propagation associated with the meridional flow speed 
at the bottom. In dynamos forced at the top with a specified period, the 
amplitude of the induced fields within the dynamo domain will be affected 
by this frequency difference. The model we present here in effect studies how 
this amplitude is affected, by treating the meridional flow at the bottom as a 
free parameter while keeping the frequency of the top forcing fixed.

In the real sun, the cycle period varies from cycle to cycle, as does the 
speed of the meridional flow and its profile with latitude. Ultimately it 
is highly desirable to include both such variations. This can be done by 
use of data assimilation techniques applied to both the surface forcing and 
meridional flow variations. As we said above, Dikpati and colleagues are 
doing that now. When that is accomplished, they may find that resonance plays 
some role. In this paper, we anticipate that possibility and focus on possible 
resonances by using a much simpler dynamo model than used in Dikpati and 
Gilman (2006), namely one that has no more than two three layers in the 
radial direction. 

Such an approach has the advantage of speed while retaining important
physical processes. But such a simple model would have little value as a tool
for prediction, because it could not be calibrated well in detail to
the sun, since it would have few degrees of freedom. It also may overestimate
the importance of resonance for the same reason. The cautions expressed in
Roald (1998) about the limits of dynamo models with one or two layers are
well taken. Nevertheless, since the forced dynamo problem has only begun to be
studied, particularly in the solar case, using a really simple model initially 
may give useful guidance about what to look for with a more realistic version. 
It is in this spirit that we report on these calculations here.

Resonance has been studied in dynamos previously, but the literature is 
small. General examples include Strauss (1986) and Reshetnyak (2010).
Resonance in the geodynamo has been studied by Stefani and Gerberth (2005)
and Fischer et al (2008). Studies for disks and galaxies include Chiba (1991),
Schmitt and R\"udiger (1992), Kuzanyan and Sokoloff (1993), and Moss (1996).
We have not located any previous studies specific to the Sun in which 
resonance has been explicitly identified and highlighted. However in all 
these areas there are almost certainly model studies in which some form 
of resonance is playing a role, but has not been brought out in the 
analysis of results. Any dynamo in which inputs such as flow fields or 
turbulent parameters such as the $\alpha$-effect are allowed to vary 
with time, either imposed or by nonlinearities internal to the system, 
such as 'quenching', could display behavior related to resonance.

\section{DYNAMO EQUATIONS AND PHYSICS}

We start from the standard flux-transport dynamo equations in vector form 
that include differential rotation, meridional circulation, turbulent magnetic
diffusivity, and allow for an inhomogeneous top boundary condition. In 
vector-invariant form, this equation is given by 

$${\partial {\B}\over \partial t} = \grad \vcross 
\left(\v \vcross \B \right) + \grad \vcross \left(\alpha \B \right) - 
\grad \vcross \left(\eta \grad \vcross \B \right) \quad\eqno(1)$$

In this equation, $\B$ is the vector magnetic field, $\v$ is the vector 
velocity, $\alpha$ is the well known alpha-effect, and $\eta$ is the 
turbulent magnetic diffusivity.

In addition, $\B$ must satisfy the divergence-free condition $\Div \B = 0$. 
This can be a problem numerically if we are solving for the three 
vector components of the magnetic field directly. But satisfaction of this 
condition is guaranteed if instead we define the magnetic field in terms of 
vector potential functions. There are at least two possible ways to do this.
The standard one from electromagnetic theory is to let $\B = \grad \vcross \A$,
'factor out' a curl operator from the whole equation. 
The resulting equation is given by

$${\partial {\A} \over \partial t} = \v \vcross \grad \vcross \A  + \alpha \grad 
\vcross \A  - \eta \grad \vcross \grad \vcross \A  \quad\eqno(2)$$

Equations (1) and (2)are quite general, including parameters and variables that
can vary with all three space dimensions, in any coordinate system. Here we 
simplify the problem to an infinite plane layer and use cartesian geometry.
We restrict the system further by assuming all quantities are independent of
one coordinate in the plane, which we take to be the $y$ coordinate. We 
identify this coordinate with longitude on the sun, so that in the cartesian
system the $x-z$ plane corresponds to the meridional plane on the sun. The 
coordinate $x$ then corresponds to colatitude and $z$ to radius. Then in
this frame, we take the velocities to be $v,u,w$ in the $x,y,z$ directions
respectively. To describe the magnetic field in this system requires only
two variables, namely the $y$ components of the toroidal field and the
poloidal potential. We denote these scalar quantities by $B$ and $A$ 
respectively. The magnetic diffusivity is denoted by $\eta$. Then the vector
system in Equations (1) and (2) reduces to a pair of scalar equations for 
$A$ and $B$ as follows

$${\partial A \over \partial t}=-v{\partial A \over \partial x}
  -w{\partial A \over \partial z} +\alpha B +\eta({\partial^2 \over \partial x^2} + {\partial^2 \over \partial z^2})A
\quad\eqno(3)$$

$${\partial B \over \partial t} = -{\partial (vB)\over \partial x}
-{\partial (wB)\over \partial z}-{\partial \over \partial x}(\alpha{\partial A \over \partial x})
  -{\partial \over \partial z}(\alpha{ \partial A \over \partial z})$$
$$  +{\partial u \over \partial z}{\partial A \over \partial x}
-{\partial u \over \partial x}{\partial A \over \partial z}
+{\partial \over \partial x}(\eta{\partial B \over \partial x})
+{\partial \over \partial z}(\eta{\partial B \over \partial z})\quad\eqno(4)$$

We simplify the problem further by restricting all coefficients in equations 
(3) and (4) to be independent of $x$. This allows for separation of variables. 
To achieve this we must require $w=0$ so that $v$ can be independent of $x$, 
and we must allow $u$ to be either a function of $x$ (latitude) only, or a 
function of $z$ (radius) only. There is evidence (Dikpati et al 2005) that 
the latitude gradient of rotation is more important than the radial gradient 
in the flux transport dynamos that best simulate solar cycles, so in this 
study we restrict ourselves to consideration of the latitude gradient. 
The diffusivity $\eta$ and the $\alpha$-effect are also taken to be independent of $x$. Equations (3) and (4) then reduce to

$${\partial \A \over \partial t}=-v{\partial A \over \partial x}
  +\alpha B   +\eta({\partial^2 \over \partial x^2} + {\partial^2 \over \partial
  z^2})A
\quad\eqno(5)$$

$${\partial B \over \partial t} = -v{\partial B\over \partial x}
-\alpha{\partial^2 A \over \partial x^2}  
  -{\partial \over \partial z}(\alpha{ \partial A \over \partial z})$$
$$  -{\partial u \over \partial x}{\partial A \over \partial z}
+\eta{\partial^2 B \over \partial x^2}
+{\partial \over \partial z}(\eta{\partial B \over \partial z})\quad\eqno(6)$$

There are a variety of boundary conditions that could be chosen for the top and 
bottom of the infinite plane layer. Consistent with solar conditions, we take
the bottom to be a perfect conductor, and therefore require $A=0$ there. $B$ in
that case is determined internally. For the top there are four plausible
alternatives: perfect conductor ($A=0$ again);insulator with no forcing ($B=0$
and $A$ matched to potential field above); forcing in potential at top ($A=A_F$)
and $B=0$ or determined internally. We make choices among these boundary
conditions when we derive the 1- and 2-layer equations in the next section.

In preparation for these derivations, we simplify the dynamo equations further 
by taking ${\partial u \over \partial x}=s,v,\alpha,\eta $ all independent of x,
and assuming all variables have solutions of the form $e^{i(kx-{\omega}t)}$ 
equations (5) and (6) reduce to

$$ -i{\omega}A=-ikvA+{\alpha}B+{\eta}(-k^2+{\partial^2 \over \partial z^2})A\quad\eqno(7)$$

$$ -i{\omega}B=-ikvB+k^2{\alpha}A-{\partial \over \partial z}({\alpha}{\partial A \over \partial z})-s{\partial A \over \partial z}-k^2B+{\partial \over \partial z}(\eta{\partial B\over \partial z})\quad\eqno(8)$$

\section{REDUCTION TO ONE, TWO AND THREE LAYER MODELS}

Before developing 1, 2 and 3-layer dynamo equations in detail, we first describe
schematically what variables and parameters are retained in these cases.
These are summarized graphically in Figure 1. For all three cases, it is 
possible to specify boundary values of the variables in addition to their 
values within each layer. We denote values at the top by a subscript $T$ 
and at the bottom by a subscript $B$. The boundary conditions shown 
correspond to a perfectly conducting bottom and an insulating or vacuum 
top with poloidal forcing $A_F$. In the 2 and 3-layer cases, we can also 
specify values of the variables at the interface between the upper and 
lower layers, namely the average of the values above and below. In some 
studies, the variables are allowed to vary with the vertical coordinate 
within the layer. Here they vary in the vertical only from layer to layer. 
Some readers may prefer the term 'level' to that of 'layer' to describe 
our model.

\begin{figure}[hbt]
\epsscale{1.0}
\plotone{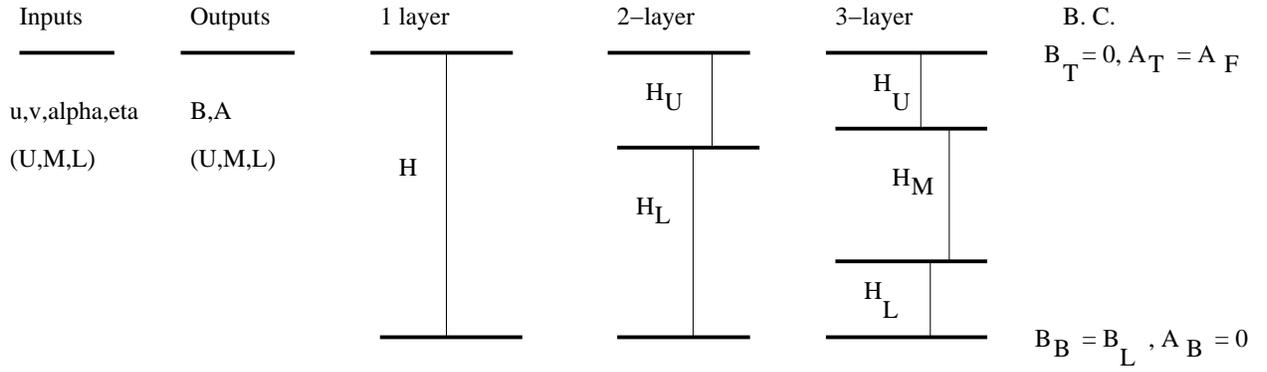}
\caption{Schematic layout of parameters and variables for one,two and three
layer models to be used. Leftmost column lists the inputs to the model
for each layer, in particular, differential rotation $u$, meridional
circulation $v$, $\alpha$-effect and magnetic diffusivity $\eta$. Next column
lists the outputs, namely toroidal field $B$ and poloidal potential $A$.
The next three columns define the layer thicknesses $H$ respectively for
one, two and three layer models. The rightmost column lists the boundary
conditions used at the top and bottom of the dynamo domain. In all cases,
the subscripts $U,M,L$ refer respectively to the upper, middle, and lower
layers.} 
\end{figure}

On the left of Figure 1 are shown the variables $B,A$ and the parameters $u,v,
\alpha,\eta$ for each layer in the model, denoted by the subscripts $U,M,L$
for three layers, $U,L$ for two layers, and unsubscripted for the 1-layer
version. The whole depth of the dynamo domain is taken to have a thickness H,
no matter how many layers it is subdivided into.

\section{1-LAYER MODEL EQUATIONS AND RESULTS}

We now use the boundary conditions defined in Figure 1 to evaluate the vertical
derivatives in equations (7) and (8). With rearrangements to put only forcing 
terms on the right hand sides, these equations become

$$[-i({\omega}-kv)+{\eta}(k^2+{8\over H^2})]A-{\alpha}B={4{\eta} \over H^2}A_F\quad\eqno(9)$$

$$-{\alpha}({8\over H^2}+k^2)A+[-i({\omega}-kv)+{\eta}(k^2+{4 \over H^2})]B=
-({4{\alpha}\over H^2}+{s\over H})A_F\quad\eqno(10)$$

We can reduce the number of parameters to take account of by making equations 
(9) and (10) dimensionless, by using $H$ as the length scale, $H^2/{\eta}$ as
the time scale and recognize that $B/A$ scales as $H$. Then velocities and 
$\alpha$ scale as ${\eta}/H$ and equations (9) and (10) become

$$[-i(\omega-kv)+k^2+8]A-{\alpha}B=4A_F\quad\eqno(11)$$

$$-{\alpha}(k^2+8)A+[-i(\omega-kv)+k^2+4]B=-(4{\alpha}+s)A_F\quad\eqno(12)$$

We find solutions to equations (11) and (12) from standard linear equation 
theory. In the case with no forcing, there are solutions only if the $(2X2)$
determinant $DT$ of the coefficients vanishes. This requires that both the real
and imaginary parts of $DT$ be zero. DT is given by  

$$DT=[-i(\omega-kv)+k^2+8][-i(\omega-kv)+k^2+4]-{\alpha}^2(k^2+8)=0,\quad\eqno(13)$$

This yields a solution for $\omega$ in terms of the mode wavenumber $k$
and the meridional flow $v$ and $\alpha$-effect, given by

$$\omega=kv-i(k^2+6\pm [(k^2+6)^2+(k^2+8)({\alpha}^2-(k^2+4)]^{1/2})\quad\eqno(14)$$

When the factor multiplying $i$ in equation (14) is real, then (14) yields
modes that propagate at the speed of the meridional flow, and either grow or
decay with time. One growing mode is assured if $k^2<{\alpha}^2-4$, so if
the latitudinal scale of the mode is sufficiently large, a growing dynamo 
mode results. This is a form of flux-transport dynamo, but one in which
differential rotation plays no role, even though it is present. This is true 
only of the one-layer model; with two or more layers, differential rotation
does play a role in the unforced dynamos.

When $k^2>{\alpha}^2-4$ then there are either all decaying modes, or
oscillatory modes that propagate at speeds different from the meridional
flow speed. We can think of the condition $k^2={\alpha}^2-4$ as the threshhold
for dynamos to occur.

In the case with forcing $A_F$ at (real) frequency ${\omega}_r$, bounded 
solutions to equations (11) and (12) exist only if the determinant $DT$ 
does not vanish. In this case, the solutions for $B$ and $A$ in terms of
$A_F$ are given by

$$B=(k^2+8+i(4{\alpha}+s)(\omega-kv))A_F/DT\quad\eqno(15)$$

$$A=(4A_F+{\alpha}B)/[-i(\omega-kv)+k^2+8]\quad\eqno(16)$$

>From equation (15) for the toroidal field $B$ we can see immediately that the
toroidal field induced in response to the poloidal forcing $A_F$ is largest 
when $DT$ is smallest, and is unbounded when $DT=0$. Therefore the system 
experiences resonance, when the frequency of the forcing equals the 'frequency' 
$kv$ associated with the meridional flow and the wavenumber of the forcing 
equals$({\alpha}^2-4)^{1/2}$. Note that this is the same point in parameter
space where there are neutral dynamo waves as found from equation (13). Just as
in resonant systems generally, here the natural frequency of the system and the
frequency of the forcing are the same. So the presence of resonance 
requires an $\alpha$-effect, but from equation (15) it does not require a 
differential rotation. The resonance in this 1-layer model is independent
of the sign of the $\alpha$-effect. It is well known that $\alpha^2$ dynamos
can have propagating unstable dynamo modes, so resonance is possible in such
systems without the effect of differential rotation. In models with two or more layers, resonance does involve differential rotation.

What is the relevance of this resonance phenomenon to the solar cycle? We know
that at the top of the convection zone, the meridional flow and surface 
poloidal forcing are propagating in opposite directions, with the poleward 
meridional flow being an order of magnitude greater. Therefore in this domain 
the dynamo should be far from resonance. But at the bottom, the meridional 
flow is toward the equator, and has a speed similar to the propagation speed of the surface poloidal source. Therefore if the surface poloidal source can be 
'felt' near the bottom of the convection zone, resonance might occur. A 
1-layer dynamo model is inadequate to test for this possibility, so we must use a model with at least two layers.

\section{2 AND 3-LAYER MODEL EQUATIONS }

\subsection{2-LAYER EQUATIONS}

For the 2-layer model equations, we proceed in the same way as with the 1-layer
equations. But here, as listed in Figure 1, most parameters have different 
values for the upper and lower layers. But for simplicity and to be able to 
achieve separation of variables, we keep the $y$ (latitudinal) gradient $s$ of
the east-west (rotational) flow the same in both layers. Then the dimensional 
2-layer equations become
 
$$[-i(\omega-kv_U)+{\eta}_U(k^2+6/{H_U}^2)]A_U-{\alpha}_UB_U-2{\eta}_UA_L/{H_U}^2=4{\eta}_UA_F/{H_U}^2\quad\eqno(17)$$

$$-({\alpha}_U(k^2+6/{H_U}^2)+s/2H_U)A_U+[-i(\omega-kv_U)+{\eta}_U(k^2+6/{H_U}^2)]B_U+(2{\alpha}_U/{H_U}^2-s/2H_U)A_L$$
$$-2{\eta}_UB_L/{H_U}^2=-(4{\alpha}_U/{H_U}^2+s/H_U)A_F\quad\eqno(18)$$

$$-2({\eta}_L/{H_L}^2)A_U+[-i(\omega-kv_L)+{\eta}_L(k^2+6/{H_L}^2)]A_L-{\alpha}_LB_L=0\quad\eqno(19)$$

$$(2{\alpha}_L/{H_L}^2+s/{2H_L})A_U-2{\eta}_L/{H_L}^2)B_U$$
$$+[-{\alpha}_L(k^2+6/{H_L}^2)+s/2H_L]A_L$$
$$+[-i(\omega-kv_L)+{\eta}_L(k^2+2/{H_L}^2)]B_L=0\quad\eqno(20)$$

By inspection, we can see that the layers are coupled by processes involving
diffusion, differential rotation and the $\alpha$-effect.
To render the system dimensionless in this case, we use $H$ for the length 
scale and $H^2/{\eta_U}$ for the time scale. We also introduce the parameters 
$P_{vL},P_{\alpha L}, P_{\eta L}$ respectively for the ratios $v_L/v_U,
{\alpha_L}/{\alpha_U},{\eta_L}/{\eta_U}$. We also define $D_U=H_U/H$ and $D_L=H_L/H$. 
Then $D_U+D_L=1$. With these definitions and evaluating the $z$ derivatives in 
equations (7) and (8), we get four equations for the toroidal field and poloidal 
potential of the two layers. In dimensionless form these equations are

$$[-i(\omega-kv_U)+k^2+6/{D_U}^2]A_U-{\alpha}_UB_U-2A_L/{D_U}^2=4A_F/{D_U}^2\quad\eqno(21)$$

$$-({\alpha}_U(k^2+6/{D_U}^2)+s/2D_U)A_U+[-i(\omega-kv_U)+k^2+6/{D_U}^2]B_U+(2{\alpha}_U/{D_U}^2-s/2D_U)A_L$$
$$-2B_L/{D_U}^2=-(4{\alpha}_U/{D_U}^2+s/D_U)A_F\quad\eqno(22)$$

$$-2(P_{\eta L}/{D_L}^2)A_U+[-i(\omega-kP_{vL}v_U)+(k^2+6/{D_L}^2)P_{\eta L}]A_L-P_{\alpha L}{\alpha}_UB_L=0\quad\eqno(23)$$

$$(2P_{\alpha L}{\alpha}_U/{D_L}^2+s/{2D_L})A_U-2(P_{\eta L}/{D_L}^2)B_U$$
$$+[-P_{\alpha L}{\alpha}_U(k^2+6/{D_L}^2)+s/2D_L]A_L$$
$$+[-i(\omega-kP_{vL}v_U)+(k^2+2/{D_L}^2)P_{\eta L}]B_L=0\quad\eqno(24)$$

By analogy with the 1-layer system, in the 2-layer case we should expect
resonance to be found near where the (4X4) determinant of the coefficients in
equations (17)-(20) vanishes. In the homogeneous case ($A_F=0$) this determinant
is a quartic equation for the complex eigenfrequency $\omega$ of the dynamos 
of the 2-layer system. Being a 4th order system, we can not in general find 
closed or simple algebraic forms for either the amplitudes of the response 
to the top forcing, or the phase speed and growth rate of the unforced dynamo. 
But only small programs are necessary to get results.

For the case with forcing, we find the amplitudes and phases of $A_U,A_L,B_U,B_L$ in terms of the amplitude and phase of $A_F$ by application of Cramers rule
(refs). We apply Cramers rule to equations (21)-(24) defined in symbolic form as

$$ c_{11}A_U+c_{12}B_U+c_{13}A_L+c_{14}B_L=F_1A_F\quad\eqno(25)$$

$$ c_{21}A_U+c_{22}B_U+c_{23}A_L+c_{24}B_L=F_2A_F\quad\eqno(26)$$

$$ c_{31}A_U+c_{32}B_U+c_{33}A_L+c_{34}B_L=0\quad\eqno(27)$$

$$ c_{41}A_U+c_{42}B_U+c_{43}A_L+c_{44}B_L=0\quad\eqno(28)$$

in which all of the $c_{i,j}$ coefficients and $F_1,F_2$ are
defined by matching terms in equations (25)-(28) with their
counterparts respectively in equations (21)-(24), so that
$c_{14}=0$ and $c_{32}=0$. Then to make Cramers rule work, 
we must have the determinant of the coefficients in equations
(25)-(28) not vanish. But where it approaches zero is where in
the parameter space we should expect resonance to occur, since 
this determinant is the denominator for the solutions of
equations (25)-(28) found by Cramers rule.

\subsection{3-LAYER EQUATIONS}

How would the results we have obtained change if we added another layer
to the system? If the results concerning resonance are similar, it gives
us more confidence that the same phenomenom may occur in much more realistic 
systems with many layers in the vertical. If not, then the results obtained
above have much more limited significance for the general case.
With three layers, the number of equations to be solved expands to six.
These are given by

$$[-i(\omega-kv_U)+k^2+{6 \over D_U^2}]A_U-\alpha_UB_U-{2 \over D_U^2}A_M={4 \over D_U^2}A_F\quad\eqno(29)$$

$$[-\alpha_U(k^2+{6 \over D_U^2})-{s \over 2D_U}]A_U+[-i(\omega-kv_U)+k^2+{6 \over D_U^2}]B_U+({2\alpha_U \over D_U^2}-{s \over 2D_U})A_M$$
$$-{2 \over D_U^2}B_M=-({4\alpha_U \over D_U^2}+{s \over D_U})A_F\quad\eqno(30)
$$

$$-{2P_{\eta M} \over D_M^2}A_U+[-i(\omega-kv_UP_{vM})+P_{\eta M}(k^2+{4 \over D_M^2})]A_M-\alpha_UP_{\alpha M}B_M-{2P_{\eta M} \over D_M^2}A_L=0\quad\eqno(31)$$

$$[{2\alpha_UP_{\alpha M} \over D_M^2}+{s \over 2D_M}]A_U-{2P_{\eta M} \over D_M^2}B_U-\alpha_UP_{\alpha M}(k^2+{4 \over  D_M^2})A_M$$
$$+[-i(\omega-kv_UP_{vM})+P_{\eta M}(k^2+{4 \over D_M^2})]B_M
+[{2\alpha_UP_{\alpha M}\over D_M^2}-{s \over 2D_M}]A_L-{2P_{\eta M} \over D_M^2}B_L=0\quad\eqno(32)$$

$$-{2P_{\eta L} \over D_L^2}A_M+[-i(\omega-kv_UP_{vL})+P_{\eta L}(k^2+{6 \over D_L^2})]A_L-\alpha_UP_{\alpha L}B_L=0\quad\eqno(33)$$

$${s \over 2D_L}A_U-{2\alpha_UP_{\alpha L} \over D_L^2}A_M-{2P_{\eta L} \over {D_L}^2}B_M+[-\alpha_UP_{\alpha L}(k^2+{6 \over D_L^2})+{s \over 2D_L}]A_L$$
$$+[-i(\omega-kv_UP_{vL})+P_{\eta L}(k^2+{2 \over {D_L}^2})]B_L=0\quad\eqno(34)$$

Similar to equations (25)-(28) above, equations (29)-(34) can be written in 
symbolic form as

$$c_{11}A_U+c_{12}B_U+c_{13}A_M+c_{14}B_M+c_{15}A_L+c_{16}B_L=F_1A_F\quad\eqno(35)$$

$$c_{21}A_U+c_{22}B_U+c_{23}A_M+c_{24}B_M+c_{25}A_L+c_{26}B_L=F_2A_F\quad\eqno(36)$$

$$c_{31}A_U+c_{32}B_U+c_{33}A_M+c_{34}B_M+c_{35}A_L+c_{36}B_L=0\quad\eqno(37)$$

$$c_{41}A_U+c_{42}B_U+c_{43}A_M+c_{44}B_M+c_{45}A_L+c_{46}B_L=0\quad\eqno(38)$$

$$c_{51}A_U+c_{52}B_U+c_{53}A_M+c_{54}B_M+c_{55}A_L+c_{56}B_L=0\quad\eqno(39)$$

$$c_{61}A_U+c_{62}B_U+c_{63}A_M+c_{64}B_M+c_{65}A_L+c_{66}B_L=0\quad\eqno(40)$$

in which, as before, the $c_{ij}$ coefficients are defined by matching terms
from equations (35)-(40) respectively with those of (29)-(34). But here there
are many more coefficients that are zero, namely $c_{14},c_{15},c_{16}; 
c_{25},c_{26}; c_{32},c_{36}; c_{51},c_{52},c_{54};c_{62}$. This greatly 
reduces the number of symbolic multiplys needed to apply Cramers rule, 
rendering it practial to work out all the algebra for this 6X6 system.

\subsection{PARAMETER CHOICES AND SCANS}

Equations (21)-(24) for the 2-layer model contain nine free parameters in 
addition to the poloidal forcing $A_F$; equations (29)-(34) for the 3-layer 
model contain 16 parameters. Therefore we must make judicious choices of 
parameter values. In making these choices we will be guided by solar conditions 
as well as the uncertainties in the solar properties that define these 
parameters.

For example, the dimensionless frequency $\omega$ of the top forcing should be
approximately the frequency of the solar cycle, corresponding to a period of
22 years. With ${\eta}_U/H^2$ our frequency scale, for ${\eta}_U=2X10^{12} 
cm^2/sec$ and $H=2X10^{10} cm$ this frequency is ~$5X10^{-9}/sec$. The solar
cycle frequency is ~$9X10^{-9}/sec$, so the dimensionless forcing frequency
should be about 1.8 units. Therefore a frequency range of 1.5 to 2. would 
cover most variability in solar cycles. In all calculations displayed below,
we have chosen a dimensionless frequency $\omega$= 1.8. Specifying the 
latitudinal wavemumber of the forcing is more uncertain. The width of the 
sunspot zone in one hemisphere is about 30 degrees latitude, or ${\pi}R/6$. 
This would be the minimum half wavelength of the forcing, but that forcing 
is seen to be broader in latitude scale than that, due to the dispersal 
and decay of active regions. Also, we never see surface fields from more than 
2 sunspot cycles at the same time, so a more reasonable wavelength might be 
${\pi}R/2$, the distance between equator and pole. Then this wavelength  
would correspond to a dimensionless wavenumber $k=1.14$ units at the surface 
and $k=1.63$ units at the depth of the tachocline. An average value would 
be about 1.4 units, which is what we use for all calculations. As for 
velocities, the velocity scale ${\eta}_U/H$ is about 1m/sec, so a 
typical solar meridional flow near the top would be 15 units, and the 
latitudinal differential rotation linear velocity relative to the rotating
frame of about $s=70$ units. 

We will use these dimensionless values to guide our choices of parameter ranges 
to survey. For some purposes the choice of ${\eta}_U=10^{12}cm^2/sec$ may be 
too high. Reducing it by a factor of ten means that all dimensionless solar
frequencies and velocities are increased by a factor of ten, but dimensionless 
wavenumbers remain the same.

\section{2 AND 3-LAYER RESULTS}

\subsection{ANALYTICAL EVIDENCE OF RESONANCE}

The 1-layer results given above could give us guidance about where to look
in parameter space for resonance when there are more layers than one. As in
that case, we might expect that for resonance to occur in a layer, we must 
have the phase speed of the forcing at the top be equal to or vary close to the 
meridional flow speed in that layer. We have established both algebraically 
and numerically that resonance does happen in the bottom most layer of the 
system when the meridional flow satisfies that condition and the cross product 
of the coefficients of $A_L$ and $B_L$ approaches zero. In terms of formulas, 
this resonance occurs in the 2-layer model in the neighborhood of

$$c_{33}c_{44}-c_{34}c_{43}=0\quad\eqno(41)$$

and in the 3-layer model near where

$$c_{55}c_{66}-c_{56}c_{65}=0\quad\eqno(42)$$

In both 2 and 3-layer models this implies

$$\omega-kP_{vL}v_U=0\quad\eqno(43)$$

and

$$P_{\eta L}^2(k^2+6/D_L^2)(k^2+2/D_L^2)-\alpha_L^2(k^2+6/D_L^2)+\alpha_Ls/2D_L=0\quad\eqno(44)$$

The fact that the conditions for resonance are identical in the 2 and 3-layer
cases suggests that by induction that this will remain true no matter how many
layers the model contains. Therefore it is likely to be a robust general 
property of this flux-transport dynamo, but we have not attempted to prove this
mathematically. It is evident from equation (44) that 
the $\alpha$-effect in the bottom layer plays an important role in creating 
resonance there. Some flux-transport models applied to the Sun contain no
$\alpha$-effect there, but Dikpati and Gilman (2001) showed that its
presence could be responsible for choosing the correct symmetry
for the Sun's toroidal and poloidal fields (see also Bonanno et al 2002,
Hotta \& Yokoyama 2010). The resonance we demonstrate in this work gives 
further importance to knowing what the $\alpha$-effect is at the base of 
the convection zone.

The conditions (43) and (44) guarantee an essentially infinitely large
resonance (interestingly even though there is diffusion in the problem,
which usually bounds the resonance to a finite value), but to be realized
requires a precise combination of values of several parameters of the 
problem, very unlikely to be realized. But just being 'close' to resonance
is enough to increase the response of the system to the same forcing
at the top by a factor of 10-100, beyond the range of variation in
solar cycle peaks. So it is worth mapping out the response over a wide
range of parameter values that are plausible for the sun. But at the same
time we must recognize that this 3-layer model is much simpler than the
real sun, and much simpler than 2D flux transport dynamos in spherical
shells.

Solutions of equation (44) in the limit of small $P_{\eta L}$ are of
particular interest, since we expect the bottom layer to have the lowest
magnetic diffusivity. In that limit the two roots are $\alpha_L=0,sD_L/12$.
In that case, there is resonance even if the $\alpha$-effect is zero in
the bottom layer.

\subsection{NUMERICAL RESULTS FOR 2 AND 3 LAYER SYSTEMS}

\subsubsection{Location of resonance in parameter space}

Here we answer the question of where in the parameter space we will find
resonance occurring. In organizing the results it is helpful to differentiate
roles played by the various parameters in the real sun, including how they
might vary with either time or space. For example, while we do not
know with precision the thickness of the bottom layer of the convection zone
where we expect the magnetic diffusivity to be small, this thickness should not
vary much with time, or probably with latitude. Therefore we can think of
the thickness as an externally specified parameter. So we would like to know
where in the range of other parameters resonance should occur, for a selection
of assumed bottom layer thicknesses. Similarly, the solar differential 
rotation does not appear to vary much with time in the sun (the well-known
torsional oscillations in rotation are about one-half of one percent of the
equatorial rotation), so we can treat it in the same way. 

The meridional flow varies more with time, but because of the density increase 
with depth, we expect the flow in the lower layer always to be much less than 
that of the upper layer and in the opposite direction. Somewhat arbitrarily we 
take the ratio of the two, $P_{v_L}=-0.1$. We also take the meridional flow of
the middle layer to equal that of the upper layer, for simplicity and with
some guidance from observations, e.g. Gizon and  Rempel (2008), that do not 
show a reversal in meridional flow at mid-depth in the solar convection zone.

>From theoretical considerations, we also do not expect the magnetic diffusivity 
in the bottom layer to vary with time, since it represents an average over many 
turbulent fluctuations; we do expect it to vary with depth, and this we have 
taken into account within the limitations of a 2 or 3 layer model. By contrast, 
the $\alpha$-effect in the bottom layer may come from the action of a 
relatively small number of global events, such as global MHD instability, so it 
might fluctuate greatly with time, perhaps even changing sign. Therefore we 
choose to display our results as plots of the $\alpha$-effect needed for 
resonance as a function of the other parameters.

\begin{figure}[hbt]
\epsscale{1.0}
\plotone{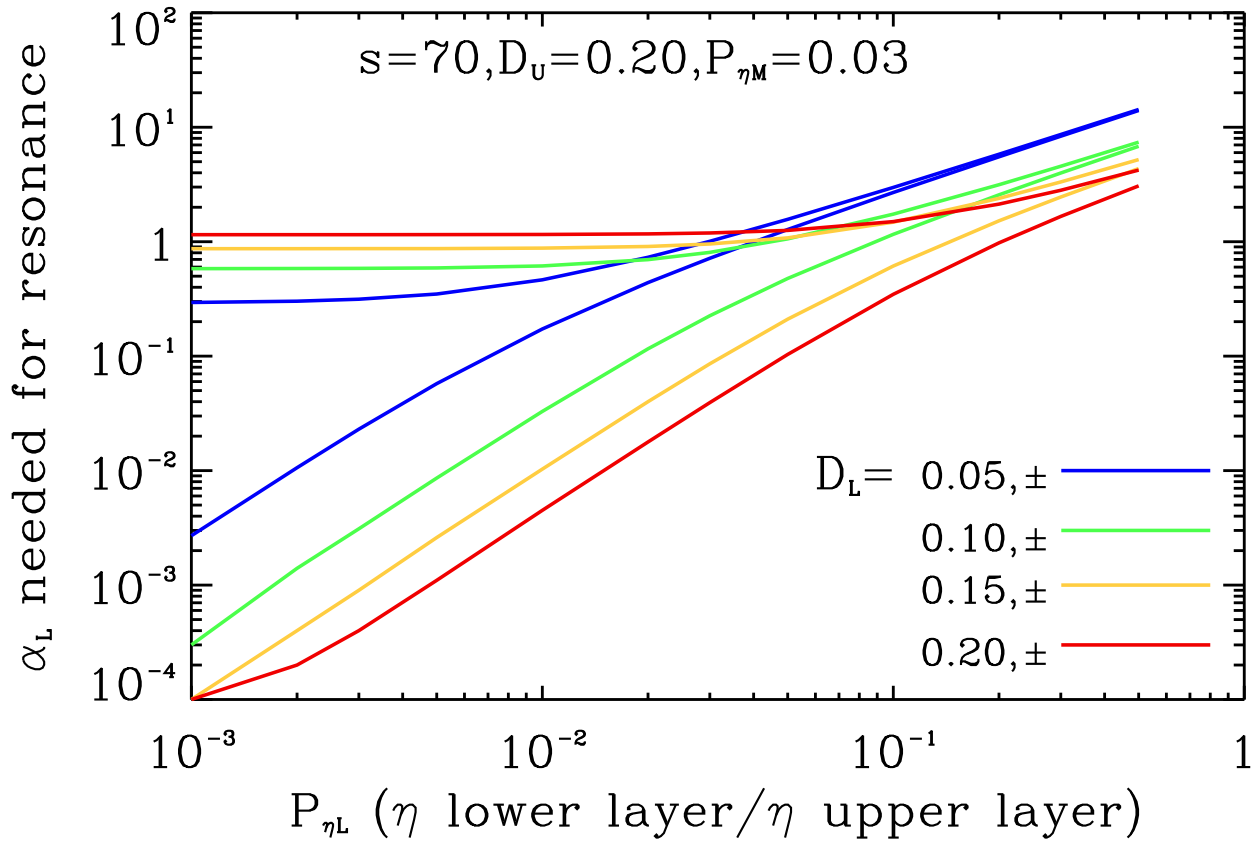}
\caption{$\alpha_L$ needed for resonance as a function of lower layer 
diffusivity. All parameters are defined in the text. There are two values
of $\alpha_L$ that give resonance for each lower layer thickness $D_L$ (two 
curves of the same color). One is positive, the other negative (sign removed 
because of log scale used). Negative values asymptote to zero as lower layer
diffusivity decreases (left hand edge of figure) while positive values 
asymptote to finite values.}
\end{figure}

Figure 2 displays $\alpha_L$ needed for resonance as a function of the 
magnetic diffusivity of the lower layer, for a solar differential rotation
($s=70$) and a selection of lower layer thicknesses (the upper layer 
thickness is held fixed at $20\% (D_U=0.2)$ of the total thickness). Since the 
y-axis is logarithmic, we have reversed the sign of all the negative $\alpha_L$ 
values. Viewed from the left hand $y$-axis, the positive $\alpha_L$ values 
are the upper family of curves, while the negative $\alpha_L$ values are the 
lower family. The negative values are all asymptotic to zero with decreasing 
$P_{\eta L}$, while the positive values are all asymptotic to constant nonzero 
values, consistent with the analytical limits shown in section 6.1. For 
increasing values of $P_{\eta L}$ the positive and negative $\alpha_L$ values 
for the same lower layer thickness are asymptotic to the same amplitudes, 
because in that limit the effect of differential rotation becomes 
insignificant. The effect of differences for different lower layer thicknesses 
also become less, because as the thickness increases,  vertical diffusion 
shrinks to an amplitude closer to the latitudinal diffusion, which is 
proportional to $k^2$, the square of the latitudinal wavenumber of the 
externally imposed forcing.

Figure 2 shows that for all lower layer magnetic diffusivities, there is always
at least one $\alpha_L \sim O(1)$ available to create resonance, and two
available if the lower layer diffusivity is high enough. This is true for 
all lower layer thicknesses shown, which cover the range of reasonable values
for the thickness of the layer with the lowest magnetic diffusivity in
the dynamo domain $(1-4X10^4km)$. A reasonable assumption for the Sun is
that the turbulent magnetic diffusivity at the bottom is a factor $10^2-10^3$
smaller than at the top. With our scaling and the curves shown in Figure 2,
this means resonance should occur for dimensional $\alpha_L$ values in the 
range $30cm/sec-1m/sec$, which are very typical values used in flux-transport
dynamos applied to the Sun.  Resonance should also occur for much
smaller values of $\alpha_L$ of the opposite sign.

We can give some physical interpretation of the results shown in Figure 2
and contained in the formulas used to generate it. There is a competition among
the physical processes associated with the $\alpha$-effect, magnetic diffusion,
differential rotation and meridional circulation to determine where in 
parameter space resonance will occur. In general, diffusion works against 
resonance, while the $\alpha$ effect and differential rotation work to produce 
resonance by inductive processes. But, depending on their signs, the 
$\alpha$-effect and differential rotation work with or against each other. From 
equation (44), if the product $\alpha_Ls$ is negative they work in concert, 
and if it is positive, they are in opposition. We have taken $s$ positive for 
the differential rotation, so when $\alpha_L<0$ they reinforce, and when 
$\alpha_L>0$ they oppose each other. 

What happens is that oppositely signed $\alpha_L$ leads to oppositely 
directed poloidal potential and therefore oppositely directed toroidal fields 
induced by the same differential rotation. All patterns are swept toward 
the 'equator' by the meridional flow. In one case the induced toroidal field 
on the leading edge of the pattern reinforces what is already present, and 
in the other case it tends to cancel it out. Clearly the former case leads 
to a stronger response, or approach to resonance. This approach to resonance is 
optimized by choosing the meridional flow speed to match the forcing speed, 
effectively 'freezing' the phase of the forcing relative to the phase of 
the induced fields, allowing for maximum amplification.

When $\alpha_L$ and $s$ are working together, for a given $s$, a smaller 
$\alpha_L$ is needed to achieve resonance. Hence for all diffusivities,
the negative $\alpha_L$ values found for resonance are smaller in amplitude
than their positive counterparts for the same values of other parameters,
as seen in Figure 2.

The differential rotation value $s=70$ in our scaling is equivalent to the
whole differential rotation of the sun with latitude at the photosphere.
But the lower layer of the model applies to the bottom of the convection
zone and the tachocline, where the latitudinal differential rotation would
be smaller by an amount that is a fairly strong function of depth. How much
difference in the values of $\alpha_L$ that are needed to produce resonance
occurs for lower differential rotation? Figures 3 and 4 give the answer.

In Figures 3 and 4 we display $\alpha_L$ needed for resonance as a function
of the differential rotation parameter $s$, for the same lower layer
thicknesses as shown in Figure 2, for two selected lower layer diffusivities 
$\eta_L=6X10^9,6X10^{10} cm^2/sec$ (Figures 3, 4 respectively),
corresponding to $P_{\eta L}=0.003,0.03$, that cover the range of plausible 
values for the bottom of the solar convection zone. Here we use a linear 
scale for $\alpha_L$ so that we can retain its sign.

\begin{figure}[hbt]
\epsscale{1.0}
\plotone{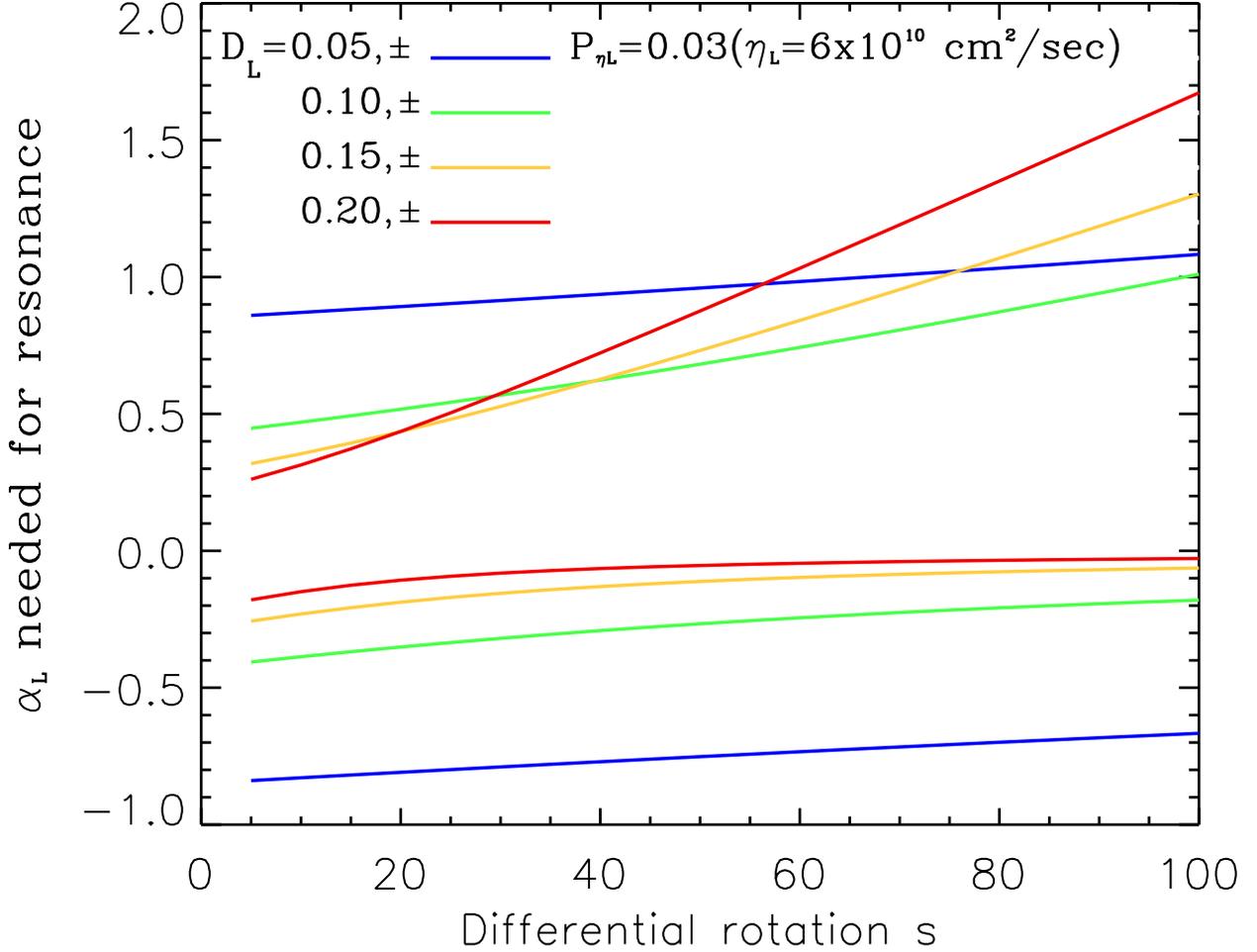}
\caption{$\alpha_L$ needed for resonance as a function of the differential 
rotation parameter $s$ for a lower layer magnetic diffusivity that is 0.03
of that of the upper layer. As in Figure 2, there are two $\alpha_L$ values,
one of each sign (two curves of same color) for each lower layer thickness
and choice of other parameters.}
\end{figure}

The primary message from Figures 3 and 4 is that, for lower layer thicknesses
and diffusivities plausible for the Sun, resonance occurs for all differential
rotations possible in the lower layer, for values of $\alpha_L \sim O(1)$.
The positive $\alpha_L$ needed for resonance increases with differential 
rotation $s$, since in this case the $\alpha$-effect and effect of differential
rotation oppose each other, while the negative $\alpha_L$ needed declines in 
amplitude with increase in $s$, because in this case the two effects 
reinforce each other.

\begin{figure}[hbt]
\epsscale{1.0}
\plotone{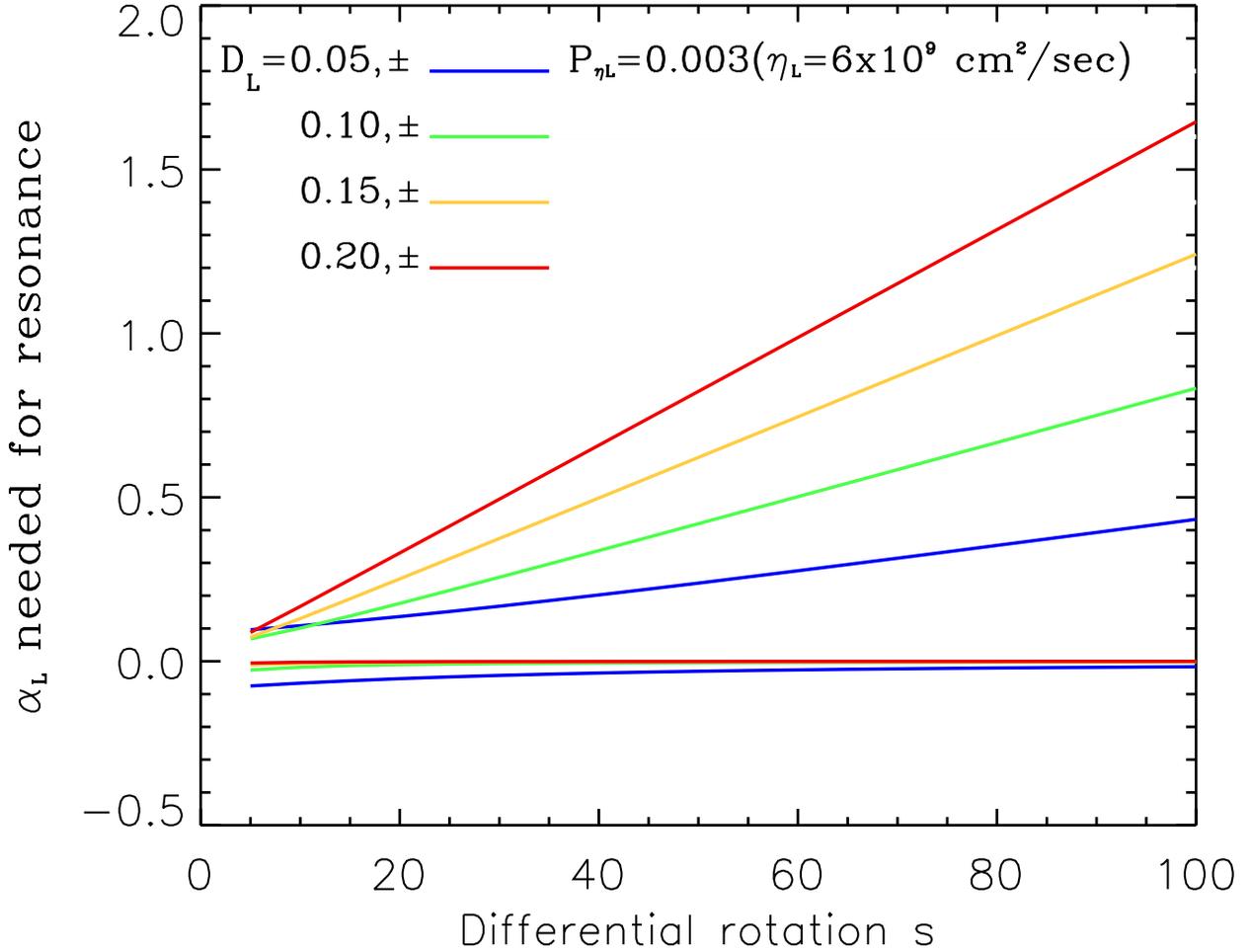}
\caption{Same as Figure 3 but for a lower layer diffusivity that is 0.003
of that of the upper layer.}
\end{figure}
The results shown in Figures 2-4 apply to both two- and three-layer  models,
since equations (43) and (44) contain quantities only from the lower layer.
But clearly the three-layer system captures more physics, so in the next 
section we focus on amplitude results for three layers. The corresponding 
results for two layers are qualitiatively similar.

\subsection{Examples of resonant response to forcing}
The previous subsection presented guidance for where in our parameter space
to find resonance. Here we present what that resonance looks like as functions
of the various parameters of the problem. Figures 5 and 6 display the 
amplitudes of toroidal field and poloidal potential for all three layers of
the model for poloidal forcing $A_F=1$ for two selected lower layer magnetic 
diffusivities, the same as for Figures 3 and 4 respectively. These results 
are for full solar differential rotation ($s=70$), a lower layer thickness of 
$10\%$ of the total thickness, and a meridional flow speed at the top of 
$v_U=-12.86$, which is predicted from equation (43) for resonance in the 
lower layer for $\omega=1.8$ and $k=1.4$ when $P_{v_L}=-0.1$. It is important

\begin{figure}[hbt]
\epsscale{1.0}
\plotone{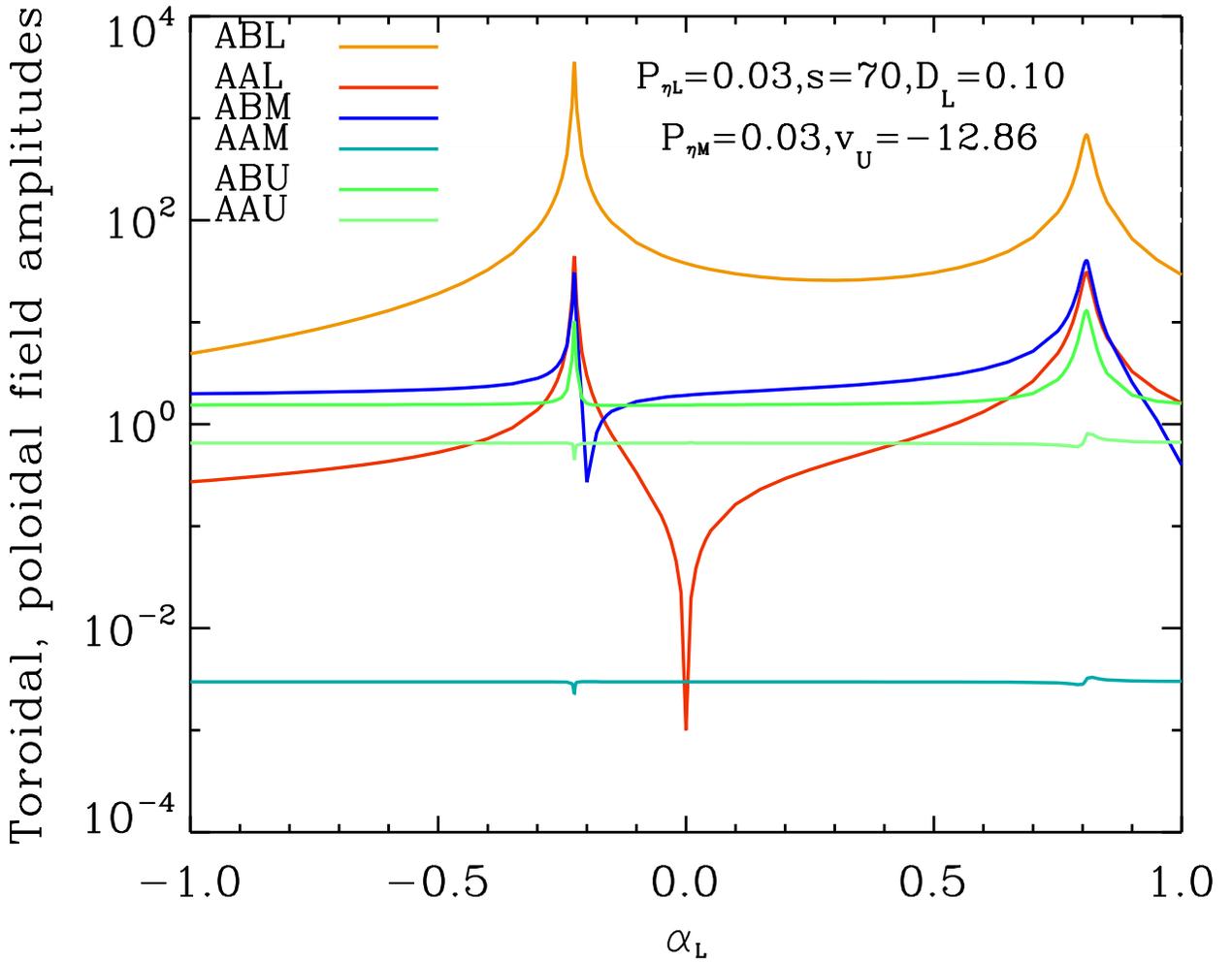}
\caption{Toroidal and poloidal field amplitudes as functions of $\alpha_L$ for 
a lower layer magnetic diffusivity of 0.03 of that of the upper layer. Color
key for all quantities shown in the upper left.}
\end{figure}

to realize that the amplitudes shown in Figures 5-7 are not exponentially
growing as in the usual unforced dynamo solutions, but instead represent
amplitudes of forced oscillatory solutions. Strictly speaking, these are not
self excited dynamos, because of the top boundary forcing, but they are 
dynamos nontheless.
\begin{figure}[hbt]
\epsscale{1.0}
\plotone{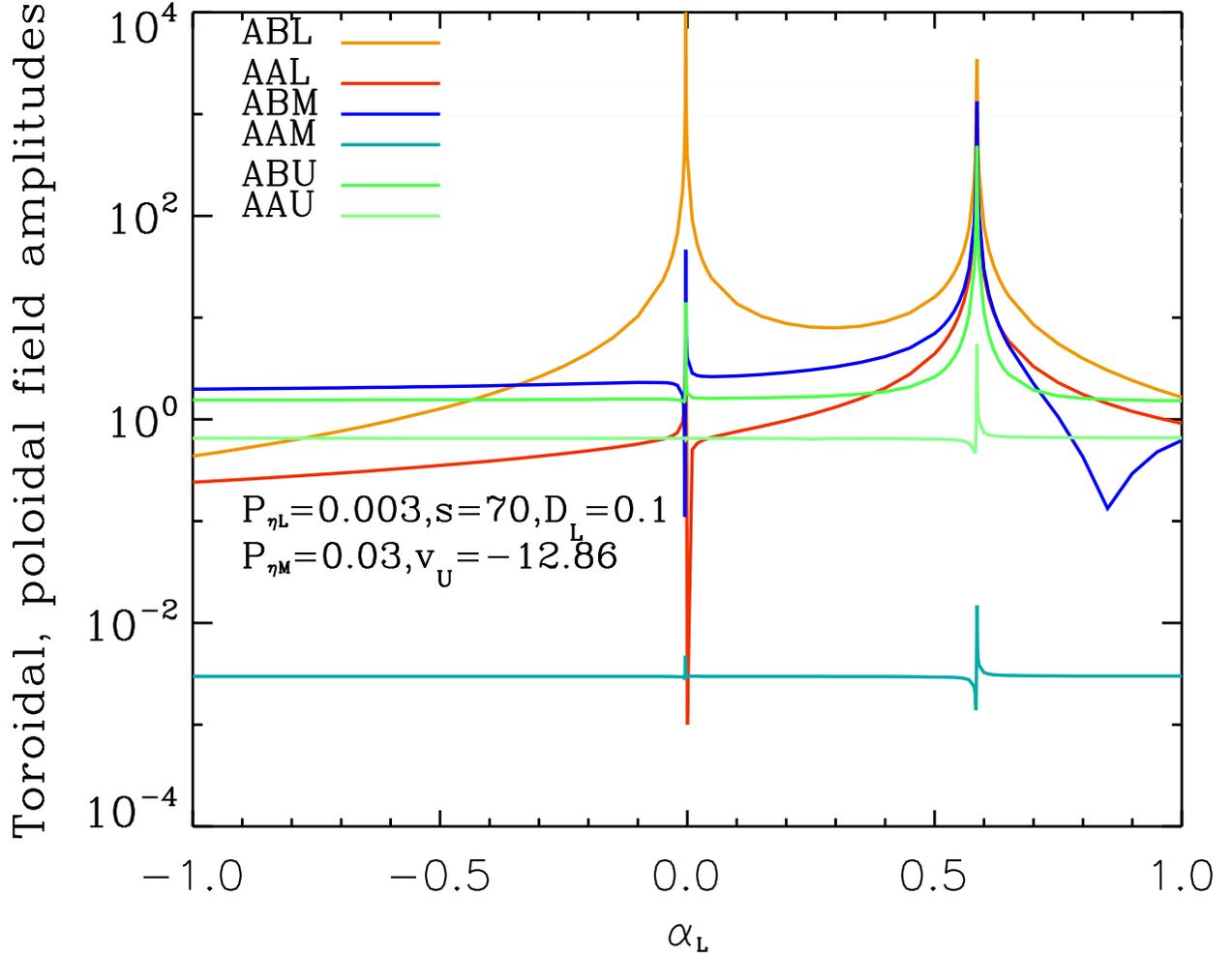}
\caption{Same as in Figure 5 but for a lower layer magnetic diffusivity of
0.003 of that of the upper layer.}
\end{figure}
For these parameter values, equations (43) and (44) predict resonance will 
occur near $\alpha_L=0.807,-0.225$ for $P_{\eta L}=0.03$ and $0.585,-0.003$ for 
$P_{\eta L}=0.003$. We see strong upward spikes in $ABL$, the amplitude 
(absolute value) of the toroidal field in the lower layer (gold curve) at about these values of $\alpha_L$. The lower layer poloidal potential $AAL$ (red curve)
also peaks there. Smaller peaks at the same $\alpha_L$ are present in the 
toroidal fields in the middle and upper layer ($ABM$, dark blue curve;$ABU$,
dark green curve) while the poloidal potentials ($AAM$, light blue curve; 
$AAU$, light green curve) of these layers respond to the resonance hardly at 
all. Finally, there is a downward spike in the poloidal potential of the 
lower layer at $\alpha_L=0$. The poloidal potential of the upper layer is 
much larger than that of the middle layer because the former is determined 
directly by the forcing at the top, while in the middle layer only a small 
$\alpha_M$ is present.

Several features of Figures 5 and 6 are notable. We see in Figure 5 that for 
all values of $\alpha_L$ there is a large response in the lower layer, 
compared to the other layers, to the forcing applied at the top, even though
the magnetic diffusivity in the lower layer is the same as that of the middle
layer. This is because the meridional flow and the $\alpha$ of the middle layer
are not close to the values needed for resonance in that layer. But at the same 
time, the middle and upper layers do show successively lower but still 
significant peaks in toroidal fields near the same values of $\alpha_L$. 
This is caused by magnetic diffusion upward across the interfaces between 
layers.

The main changes seen in Figure 6 compared to Figure 5 are that with
diffusivity of the lower layer reduced by a factor of ten, the resonance 
becomes much narrower and sharper. In other words, with lower diffusivity
the range of $\alpha_L$ over which there is substantial amplification of
the effect of the forcing at the top is narrower. But where resonance does 
occur, the middle and upper layers respond more strongly to the resonance
for non-zero $\alpha_L$. This is not true for the resonance near $\alpha_L=0$,
because such a low value leads to less production of poloidal field in the 
lower layer (compare the red curves in Figures 5 and 6 in the neighborhood of
the resonance for negative $\alpha_L$), from which the lower layer toroidal 
field must be produced by the differential rotation there.

Figures 5 and 6 are for a rather precisely chosen lower layer meridional flow
speed. What happens to the resonance if we move away from that speed? We
have examined this question by computing amplitudes of toroidal and poloidal
fields as functions of $\alpha_L$ for other speeds, namely $v_U=-15,-14,-13,
-12,-11$. We find the same resonances as seen in Figures 5 and 6, but with
somewhat different amplification factors. Thus the presence of resonance of
some significant amplitude is not strongly dependent on the precise value of
the meridional flow. This means in effect that equation (44) more closely 
determines the resonance than does equation (43). In keeping with this 
inference, if we choose a value of $\alpha_L$ only a few percent away from that
predicted to give resonance, the resonance practically disappears no matter 
what speed of meridional flow is taken.

Figure 7 displays the amplitude of the induced toroidal field in the lower
layer as a function of meridional flow speed $v_U$ in the upper layer, for 
the $\alpha_L$ values predicted for resonance for the same parameter choices as 
in Figures 5 and 6. We see that for $P_{\eta L}=.03$, the peak field does not 
occur at $v_U=-12.86$, at which equation (43) is satisfied, but rather at 
values above and below that (blue and green curves, respectively), depending 
on the $\alpha_L$ chosen. While resonance still occurs for the predicted 
$\alpha_L$ values, it is a factor of five to ten smaller than at the peaks 
shown. This is again evidence that the closeness of equation (44) to being 
satisfied is the determining factor in closeness to resonance. This is what 
makes the denominator of the algebraic expressions for $ABL$ and $AAL$ 
smallest. 

\begin{figure}[hbt]
\epsscale{1.0}
\plotone{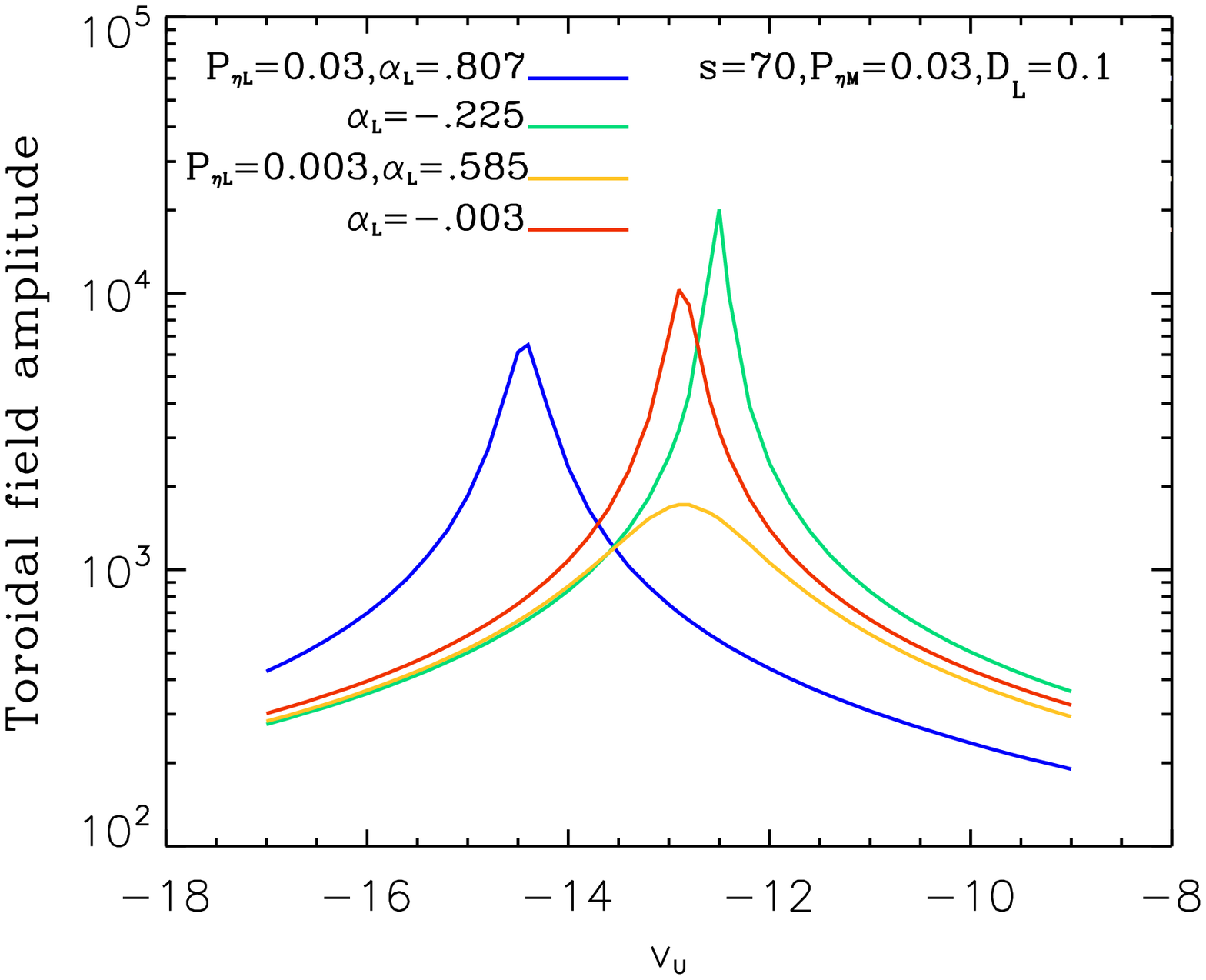}
\caption{Toroidal field amplitude of the lower layer as a function of 
the meridional flow amplitude, for four $\alpha_L$ values (see color
key in upper left of figure) at which equations (43) and (44) predict 
resonance should occur.}
\end{figure}
This departure from the predicted points of resonance occurs because 
of the amplitude of the diffusivity assumed for the lower layer. This is 
evident from Figure 7 because the peak toroidal fields produced for 
$P_{\eta L}=0.003$ (red and yellow curves) occur very close to the $\alpha_L$ 
values predicted for resonance. This also means that the resonant peaks shown
in Figure 5 are not the highest that are possible for $P_{\alpha_L}=0.03$, but
the resonance is still quite pronounced for the parameter values chosen 
for Figure 5.

\section{CONCLUSIONS AND DISCUSSION }

Our principal conclusion is that, at least in this simple
dynamo model, resonance will always be found, provided
the right parameter values are chosen. Furthermore, it occurs
for parameter choices that are plausible for the sun. It occurs
in the lowest layer of the model, where the meridional flow
toward the equator is most likely to match, or nearly match,
the propagation speed of the forcing imposed at the top, 
corresponding to the photosphere on the sun. And the effect is
large--a factor of 10-100 amplification of toroidal field
compared to the case where the parameter values are far from the
ones for which resonance is predicted. We acknowledge that this large
a difference in amplitude for different parameter values is partly
due to the model being kinematic. If $jXB$ forces were included,
the peaks would almost surely be smaller.

The solar convection zone is a spherical shell, not an infinite 
layer, so how well should we expect our results to apply to the Sun? 
In our model, the differential rotation $s$ is a constant, so the rotation 
is a linear function of 'latitude' only, and independent of depth. 
In the Sun, differential rotation varies with both latitude and depth.
In our model, meridional flow is independent of latitude,
and there is no vertical motion, while in the Sun, meridional
flow is a closed circulation confined by the boundaries of the 
convection zone as well as the poles. Meridional circulation is also known 
to vary with time on a variety of timescales. The $\alpha$-effect will also 
be a function of latitude as well as time.

These differences between our model and the Sun imply that resonance, 
if it occurs, would likely be found in more localized regions of 
the dynamo domain. But in principle it could still occur. If a 
particular location experiences resonance or near resonance for 
some period of time, then the toroidal field might amplify quickly 
there, leading to locally anomalously large field. How quickly presumably
depends on the nearness to conditions for resonance, and how long these
conditions last. Could such a sudden amplification be a precursor for 
the formation of a rising flux tube that creates a new active region? 
This seems like a hypothesis worth testing in the future.

It is also possible that in the convection zone conditions
for resonance could persist for large fractions of a solar cycle and
perhaps even from one cycle to the next. If this happens, it
could contribute to the evolution of the envelope of cycle amplitudes.
On extremely long time scales could persistent conditions 
favoring resonance be responsible for 'grand maxima', and 
conditions unfavorable for resonance lead to 'grand minima',
such as the Maunder minimum? Clearly these are very speculative
questions, requiring much more sophisticated solar dynamos 
than we have used here, to answer.

In flux-transport dynamos in spherical shells that are forced
from the top, it is the radial flow of the closed meridional
circulation that transports the forcing signal including its
frequency to the bottom. In our infinite plane cartesian model,
there is no radial flow, so the signal should be thought of as 
getting to the bottom locally via vertical diffusion and induction. 
We can think of the meridional circulation in the upper layer closing 
at infinity and returning in the lower layer.  

Since we know that not all sunspot cycles have the same duration, 
the meridional flow will be bringing a frequency to the bottom 
that could be different than that implied by the meridional 
circulation there at the time the active regions were formed, 
that gave rise to the forcing frequency. Also, a change in the
speed of the flow carrying the forcing signal will change the
frequency of that signal by Doppler-shifting it. But in percentage terms 
the differences will not be large, so there could be continually 
evolving 'nearness' to conditions needed for resonance, creating 
continuous changes in cycle amplitude, in addition to the usual 
evolution of a sunspot cyle through ascending, peak, declining 
and minimum phases. Again, investigating such possibilities 
requires much more realistic dynamo models than we have used here.

THe National Center for Atmospheric Research is sponsored by the National
Science Foundation. This work is partially supported by the NASA Living
with a Star Program through Award NNX08AQ34G.

\end{document}